\documentclass[graybox]{svmult}
\usepackage{mathptmx}       
\usepackage{helvet}         
\usepackage{courier}        
\usepackage{type1cm}        
\usepackage{makeidx}         
\usepackage{graphicx}        
\usepackage{multicol}        
\usepackage[bottom]{footmisc}
\usepackage{url}
\usepackage{amsmath,amssymb,cite}
\makeindex             

\begin{document}
\title*{Black Hole Thermodynamics\\ and Statistical Mechanics}
\author{Steven Carlip}
\institute{Steven Carlip \at Physics Department, University of 
  California at Davis, Davis, CA 95616, USA, and ITF, Utrecht 
  University, 3584 CE Utrecht, the Netherlands,
  \email{carlip@physics.ucdavis.edu}}
\maketitle

\abstract{We have known for more than thirty years that black holes
behave as thermodynamic systems, radiating as black bodies with 
characteristic temperatures and entropies.  This behavior is not only 
interesting in its own right; it could also, through a statistical
mechanical description, cast light on some of the deep problems of 
quantizing gravity.  In these lectures, I review what we currently 
know about black hole thermodynamics and statistical mechanics, 
suggest a rather speculative ``universal'' characterization of the 
underlying states, and describe some key open questions.}

\section{Introduction}

Black holes are black bodies.  

Since the seminal work of Hawking 
\cite{carHawking} and Bekenstein \cite{carBekenstein}, we have 
understood that black holes behave as thermodynamic objects, with 
characteristic temperatures and entropies.  Hawking radiation has 
not yet been directly observed, of course; a typical stellar mass 
black hole has a Hawking temperature of well under a microkelvin, 
far lower than that of the cosmic microwave background.  But the 
thermodynamic properties of black holes are well understood, having 
been been confirmed by a great many independent methods that all 
yield the same quantitative results: a 
\index{Hawking temperature}%
temperature 
\begin{equation}
kT_{\scriptscriptstyle\mathit Hawking} = \frac{\hbar\kappa}{2\pi} 
\label{carBHradiate3}
\end{equation}
and an
\index{Bekenstein-Hawking entropy}%
\index{entropy!black hole}%
entropy
\begin{equation}
S_{\scriptscriptstyle\mathit BH} 
 = \frac{\ A_{\mathit\scriptstyle horizon}}{4\hbar G} ,
\label{carintro1}
\end{equation}
where $A_{\mathit\scriptstyle horizon}$ is the horizon area and
$\kappa$ is the surface gravity.

In a typical thermodynamic system, thermal properties are 
the macroscopic echoes of microscopic physics.  Temperature 
is a measure of the average energy of microscopic constituents; 
entropy counts the number of microstates.  It is natural to 
ask whether the same is true for the black hole.  This is 
an important question: the Bekenstein-Hawking entropy 
depends on both Planck's 
and Newton's constants, and a statistical mechanical description 
of black hole thermodynamics might tell us something profound 
about quantum gravity.  Until about ten years ago, virtually 
nothing was known about black hole statistical mechanics.  Today, 
in contrast, we suffer an embarrassment of riches: we have many 
competing microscopic pictures, describing different states and
different dynamics but all predicting the same thermodynamic 
behavior.

In these lectures, I will review what is currently know---and 
not known---about black hole thermodynamics and statistical 
mechanics.  This is a large subject, and I will have to skip 
many interesting aspects.  In particular, I will not discuss 
stability analysis, the peculiarities of negative heat capacity, 
or the complicated question of black hole phase transitions,
and I will only lightly touch upon the profound issues of
information loss and holography.  

Even so, my approach will necessarily be sketchy and 
idiosyncratic, though I will also try to suggest further 
references with different emphases and different degrees of 
detail.  I will aim for a broad overview, rather than focusing 
on the fine points of any one particular approach.  Some books 
and review articles that I have found helpful include 
\cite{carWald,carJacobson,carWaldbk,carFrolov,carLesHouches}.  
In an appendix, I discuss basic black hole properties and 
explain my notation.

\section{Black Hole Thermodynamics}

I will begin with two somewhat intuitive routes to black hole 
thermodynamics.  One of these is based on the second law of 
thermodynamics, the other on the four laws of black hole mechanics.  
Neither route is completely convincing, but together they provide a 
good foundation for some of the harder quantitative approaches that 
I shall discuss later.

\subsection{Entropy and the second law \label{carsecEntropy}}

\index{Bekenstein-Hawking entropy}%
\index{entropy!black hole}%
\index{generalized second law of thermodynamics}%
\index{second law of thermodynamics}%
Imagine dropping a small box of hot gas into a black hole.  The 
initial state includes the gas and the black hole; the final state 
consists solely of a slightly larger black hole.  The initial 
state certainly has nonzero entropy, in the form of the entropy 
of the gas.  If the second law of thermodynamics is to hold, the 
final state must have nonzero entropy as well: the larger black 
hole must gain enough entropy to compensate for the entropy lost 
when the gas disappears behind the horizon.

We can make this argument somewhat more quantitative 
\cite{carKieferb}.  Suppose the box of gas has linear size $L$, mass 
$m$, and temperature $T$, and that the black hole has mass $M$ and 
horizon radius $R=2GM$ (and thus horizon area $A = 16\pi G^2M^2$).  
The box of gas will merge with the black hole when its proper 
distance $\rho$ from the horizon is of order $L$, at which point 
the disappearance of the gas will lead to a loss of entropy
$$\Delta S \sim -m/T .$$
For a Schwarzschild black hole, the proper distance from the horizon 
is
$$\rho = \int_{2GM}^{2GM + \delta r} \frac{dr}{\sqrt{1- 2GM/r}}
 \sim \sqrt{GM\delta r} ,$$
so $\rho \sim L$ when $\delta r \sim L^2/GM$.  The gas initially 
has mass $m$, but its energy as seen from infinity is red shifted 
as the box falls toward the black hole; when the box reaches 
$r = 2GM + \delta r$, the black hole will gain a mass 
$$\Delta M \sim m \sqrt{1 - \frac{2GM}{2GM + \delta r}} 
 \sim \frac{mL}{GM}.$$
If we now suppose that the box must be as large as the thermal 
wavelength of the gas, $L\sim \hbar/T$, we see that
$$\Delta S \sim -\frac{mL}{\hbar} \sim -\frac{GM\Delta M}{\hbar} 
  \sim -\frac{\Delta A}{\hbar G} .$$
To preserve the second law of thermodynamics, the black hole must 
gain an entropy of at least order $\Delta A/\hbar G$.

One can perform a similar analysis for a single particle falling  
into a Kerr black hole (assuming the particle contains at least 
one bit of entropy) \cite{carBekenstein}, a box containing a 
simple harmonic oscillator \cite{carBekenstein}, and, using a 
more sophisticated analysis, a much more general system falling 
through a horizon \cite{carWald,carBekensteinb,carZurek,carPage}.  
In each case, a ``generalized second law'' holds, provided one 
includes a change of entropy of order $\Delta A/\hbar G$ for the 
black hole.  Such reasoning led Bekenstein to suggest in 1972 
that a black hole should itself be attributed an entropy of order 
$A/\hbar G$ \cite{carBekenstein}.
 
At the time, there seemed to be a compelling argument against such 
a hypothesis.  Classical black holes are, after all, black:  
when placed in contact with a heat bath they will absorb energy
while emitting none, thus behaving as if they have a temperature 
of zero \cite{carBardeen}.  Two years later, Hawking showed that 
this problem was cured by quantum theory.  I shall return to 
this result below, but let us first consider another classical 
argument for black hole thermodynamics.

\subsection{The four laws of black hole mechanics 
 \label{carFourLaws}}

\index{laws of black hole mechanics|see {black hole mechanics,%
 laws of}}%
\index{black hole mechanics!laws of}%
In four spacetime dimensions, a stationary asymptotically flat black 
hole is uniquely characterized by its mass $M$, angular momentum 
$J$, and charge $Q$.  (In the presence of nonabelian gauge 
fields or certain exotic scalar fields, other kinds of black hole 
``hair'' can occur \cite{carWinstanley}, but this does not change 
the basic argument.)  In the early 1970s, a set of relations among 
neighboring solutions were found, culminating in Bardeen, Carter, and 
Hawking's ``four laws of black hole mechanics'' \cite{carLesHouches,%
carBardeen}.  These take a form strikingly similar to the four laws 
of thermodynamics:

\begin{enumerate}{\addtocounter{enumi}{-1}}
\item The surface gravity $\kappa$ is constant over the event horizon.
\item For any two stationary black holes differing only by small  
variations in the parameters $M$, $J$, and $Q$,  
\begin{equation}
\delta M 
 = \frac{\kappa}{8\pi G}\delta A + \Omega_H\delta J + \Phi_H\delta Q ,
\label{carFourLaws1}
\end{equation}
where $\Omega_H$ is the angular velocity and $\Phi_H$ is the electric 
potential at the horizon.
\item The area of the event horizon of a black hole never decreases,
$$ \delta A \ge 0 .$$
\item It is impossible by any procedure to reduce the surface gravity
$\kappa$ to zero in a finite number of steps.
\end{enumerate}

As in ordinary thermodynamics, there are a number of formulations of
the third law, which are not strictly equivalent; for a proof of the
version given here, which is analogous to the Nernst form of the 
third law of thermodynamics, see \cite{carIsrael}.  These laws can 
be generalized beyond the particular four-dimensional ``electrovac'' 
setting in which they were first formulated; 
\index{isolated horizon}%
the first law, in particular, holds for 
arbitrary isolated horizons \cite{carAshtekar}, and for much 
more general gravitational actions, for which the entropy can be 
understood as a Noether charge \cite{carWaldc}.

Bardeen, Carter, and Hawking noted that these laws closely parallel 
the ordinary laws of thermodynamics, with the horizon area playing 
the role of entropy and the surface gravity playing the role of 
temperature.  But they added, ``It should however be emphasized 
that $\kappa/8\pi$ and $A$ are distinct from the temperature and 
entropy of the black hole.  In fact the effective temperature of a 
black hole is absolute zero.\dots In this sense a black hole can be 
said to transcend the second law of thermodynamics.''\footnote{See 
\cite{carBardeen}, p.\ 168.}

\subsection{Black holes radiate \label{carBHradiate}}

The first suggestion that black holes might emit radiation was made
by Zel'dovich \cite{carZeldovich}, but his argument was qualitative,
and applied only to superradiant modes of rotating black holes.  
\index{Hawking radiation}%
In 1974, though, Hawking demonstrated that all black holes emit 
blackbody radiation \cite{carHawking,carHawkingb}.  The result was 
startling, and according to Page \cite{carPageb}, Hawking himself 
did not initially believe it.  In hindsight, though, one can give 
a somewhat intuitive description of the effect \cite{carSchutz}.

Such a description has two main ingredients.  The first is that 
the quantum mechanical vacuum is filled with virtual particle-%
antiparticle pairs that fluctuate briefly into and out of existence.  
Energy is conserved, so one member of each pair must have negative 
energy.  (To avoid a common confusion, note that either the particle 
or the antiparticle can be the negative-energy partner.)  Normally, 
negative energy is forbidden---in a stable quantum field theory, the 
vacuum must be the lowest energy state---but energy has a quantum 
mechanical uncertainty of order $\hbar/t$, so a virtual pair of 
energy $\pm E$ can exist for a time of order $\hbar/E$.  The 
existence of such virtual pairs is experimentally well-tested: 
for example, virtual pairs of charged particles make the vacuum a 
polarizable medium, and vacuum polarization is observed in such 
phenomena as the Lamb shift and in energy levels of muonic atoms 
\cite{carWeinberg}.

The second ingredient is the observation that in general relativity,
energy---and, in particular, the sign of energy---can be frame-%
dependent.  The easiest way to see this is to note that the 
Hamiltonian is the generator of time translations, and thus
depends on one's choice of a time coordinate.  One must therefore 
be careful about what one means by positive and negative energy
for a virtual pair.

In particular, consider the Schwarzschild metric,
\begin{equation}
ds^2 = \left(1 - \frac{2GM}{r}\right)dt^2 
   - \left(1 - \frac{2GM}{r}\right)^{-1}dr^2 - r^2d\Omega^2 .
\label{carBHradiate1}
\end{equation}
Outside the event horizon, $t$ is the usual time coordinate, 
measuring the proper time of an observer at infinity.  Inside 
the horizon, though, components of the metric change sign, 
and $r$ becomes a time coordinate, while $t$ becomes a spatial 
coordinate: an observer moving forward in time is one moving in 
the direction of decreasing $r$, and not necessarily increasing 
$t$.\footnote{Strictly speaking, the coordinates labeled $r$ 
and $t$ for $r>2GM$ are different from those with the same labels 
for $r<2GM$, since the Schwarzschild coordinate system is only 
defined in nonoverlapping patches inside and outside the horizon.  
But one can rephrase the argument in terms of proper time of 
infalling observers in a way that dodges this mathematical 
subtlety \cite{carSchutz}.}  Hence an ingoing virtual particle 
that has negative energy relative to an external observer may 
have positive energy relative to an observer inside the horizon. 
The uncertainty principle can thus be circumvented: if the 
negative-energy member of a virtual pair crosses the horizon, 
it need no longer vanish in a time $\hbar/E$, and its 
positive-energy partner may escape to infinity.

We can again make this argument a bit more quantitative.  Consider a 
virtual pair momentarily at rest at a coordinate distance $\delta r$ 
from the horizon.  As in section \ref{carsecEntropy}, the proper time 
for one member of the pair to reach the horizon will be
$$\tau \sim \sqrt{GM\delta r} .$$
Setting this equal to the lifetime $\hbar/E$ of the pair, we find 
that
$$|E| \sim \frac{\hbar}{\sqrt{GM\delta r}}, $$
which should also be the energy of the escaping positive-energy 
partner.  This is the energy at $2GM + \delta r$, though; the energy
at infinity will be red shifted to
\begin{equation}
E_\infty \sim 
  \frac{\hbar}{\sqrt{GM\delta r}}\sqrt{1 - \frac{2GM}{2GM+\delta r}}
  \sim \frac{\hbar}{GM} ,
\label{carBHradiate2}
\end{equation}
independent of the initial position $\delta r$.  We might thus 
expect a black hole to radiate with a characteristic temperature 
$kT\sim\hbar/GM$.  In fact, the precise computations I shall describe 
below yield a temperature $kT_{\scriptscriptstyle\mathit Hawking} 
= \hbar\kappa/2\pi$, which for a Schwarzschild black hole is 
$\hbar/8\pi GM$.

Inserting the Hawking temperature (\ref{carBHradiate3}) into the 
first law of black hole mechanics (\ref{carFourLaws1}), we see 
that black holes can indeed be viewed as thermal objects, with 
an entropy (\ref{carintro1}).  This result is fundamentally 
quantum mechanical---the Hawking temperature depends explicitly 
on $\hbar$---and in some sense quantum gravitational, since the 
Bekenstein-Hawking entropy depends on $G$ as well.

\subsection{Can Hawking radiation be observed?}

\index{Hawking radiation!observation}%
I will return to the more precise and detailed derivations of 
Hawking radiation below.  But let us first address the question 
of whether this effect can be observed.

For a black hole of mass $M$, the 
\index{Hawking temperature}%
Hawking temperature (\ref{carBHradiate2}) is  
$$T_{\scriptscriptstyle\mathit Hawking} 
  \sim 6\times10^{-8}\left(\frac{M_\odot}{M}\right)\, K,$$
some eight orders of magnitude smaller than the cosmic microwave 
background temperature for a stellar mass black hole and far 
smaller for a supermassive black hole.  While there is a chance 
that we could see Hawking radiation from the final stages of 
evaporation of primordial black holes \cite{carMacGibbon,carCline}, 
such events are expected to be rare and difficult to identify.

\index{TeV-scale gravity}%
\index{gravity!TeV-scale}%
Another highly speculative possibility for the detection of 
Hawking radiation comes from models of TeV-scale gravity.  In 
such models---which typically arise from ``brane world'' 
scenarios in which our four-dimensional universe is a submanifold 
of a higher-dimensional spacetime---gravity may become strong at 
energies far below the Planck scale.  If this is the case, black 
holes might be produced copiously at accelerators such as the 
LHC, and their quantum properties could be studied in detail 
\cite{carGiddings,carKanti}. 

\index{analog gravity}%
\index{gravity!analog}%
\index{dumb holes}%
A third, less direct, route is to look for analogs of Hawking 
radiation in condensed matter systems.  As Unruh first pointed out 
\cite{carUnruh}, one can create a sonic event horizon in a fluid 
flow by allowing the flow to become supersonic beyond some boundary.  
The same analysis that predicts Hawking radiation from a black hole 
leads to a prediction of phonon radiation from the sonic horizon of 
such a ``dumb hole.''  Similar phenomena can occur in a variety of 
condensed matter systems, from Bose-Einstein condensates to ``slow 
light'' to superfluid quasiparticles, and a number of experimental 
efforts are underway; for reviews, see \cite{carBarcelo,carNovello}.  
It is worth emphasizing that while such experiments could provide  
strong evidence for Hawking radiation, which is essentially a 
kinematical property, they would not test the Bekenstein-Hawking 
entropy, which depends critically on the dynamics of general 
relativity \cite{carVisser}.

\subsection{The many derivations of Hawking radiation 
\label{carManyT}}

In the absence of direct experimental evidence, how confident should 
we be about Hawking radiation and black hole thermodynamics?  
Although Hawking's derivation involves only standard quantum field 
theory, we can see from the arguments of section \ref{carBHradiate} 
that the radiation involves modes with arbitrarily high energies: 
while the asymptotic energies (\ref{carBHradiate2}) may be small, 
they come from red-shifted quanta with much higher energies near 
the horizon.  This has led some to suggest that the derivations 
might involve an extrapolation of quantum field theory beyond the 
range it can be trusted \cite{carUnruh,carJacobsonb,carHelfer}.

I will return this issue below, but for now let me suggest a
partial answer.  If only one derivation of Hawking radiation 
existed, we would clearly need to look very carefully for hidden 
assumptions and unjustified extrapolations.  In fact, though, we 
have a rather large number of different derivations, which involve 
very different assumptions and extrapolations and nevertheless all 
agree.  Some of these derivations look at eternal black holes, 
others at black holes formed from collapse; some involve explicit, 
detailed computations in particular field theories, others use 
general properties of axiomatic quantum field theory; some involve 
Planck-scale fluctuations, others cut off energies well below the 
Planck scale; some some predict only the Hawking temperature, others 
also allow a computation of the Bekenstein-Hawking entropy.  While 
it is still possible that these derivations all share a common 
flawed assumption, it seems unlikely that so many methods would 
converge on the same answer if that answer were wrong.  None of 
this vitiates the need for observational tests---after all, the 
entire general relativistic description of black holes could be 
wrong---but it suggests that a failure of black hole thermodynamics
would have to be either very subtle or very radical.

I will describe some of these derivations below.  Given the nature 
of these lectures, I will not attempt a full description of any one 
method; my aim is to give a broad overview, with references that 
will allow the reader to delve into individual approaches in more 
detail.

\subsubsection{Bogoliubov transformations and inequivalent vacua
\label{carBogol}}

\index{Bogoliubov transformation}%
As noted above, a crucial ingredient in understanding Hawking 
radiation is the fact that energy---and, in particular, ``positive'' 
and ``negative'' energy---is frame-dependent.  Consider, for 
simplicity, a free real scalar field $\varphi$.  Recall that in 
ordinary quantum field theory in flat spacetime, we quantize 
$\varphi$ by first decomposing the field into Fourier modes,
\begin{equation}
\varphi = \sum_{\bf k} \left(a_{\bf k}u_{\bf k}(t,{\bf x}) 
  + a_{\bf k}^\dagger u_{\bf k}^*(t,{\bf x})\right) \ \ \ 
  \mathrm{with}
  \ \ u_{\bf k} = e^{i{\bf k}\cdot{\bf x} - i\omega_{\bf k}t},\ \ 
  \omega_{\bf k} = \left(|{\bf k}|^2 + m^2\right)^{1/2} ,
\label{carBog1}
\end{equation}
and then interpret the $a_{\bf k}$ as annihilation operators and 
the $a_{\bf k}^\dagger$ as creation operators.  The Fourier modes 
$u_{\bf k}$ can be understood as a set of orthonormal functions 
satisfying
\begin{equation} 
(\Box + m^2) u_{\bf k}(t,{\bf x}) = 0, \qquad 
 \partial_t u_{\bf k}(t,{\bf x}) 
   = -i\omega_{\bf k}u_{\bf k}(t,{\bf x}), 
\label{carBog2}
\end{equation}
where the second condition determines what we mean by positive and 
negative frequency, and thus allows us to distinguish creation and 
annihilation operators.  The vacuum is then defined as the state 
annihilated by all of the $a_{\bf k}$,
$$a_{\bf k} |0\rangle = 0 .$$

In a curved spacetime, or a noninertial coordinate system in flat 
spacetime, standard Fourier modes are no longer available.  With a 
choice of time coordinate $t$, though, one can still find modes of 
the form (\ref{carBog2}) and perform a decomposition (\ref{carBog1}) 
to obtain creation and annihilation operators.  Given two different
reference frames with time coordinates $t$ and ${\bar t}$, two such 
decompositions exist:
\begin{equation}
\varphi = \sum_i \left(a_iu_i + a_i^\dagger u_i^*\right)  
  = \sum_i\left({\bar a}_i{\bar u}_i 
    + {\bar a}_i^\dagger{\bar u}_i^*\right) ,
\label{carBog3}
\end{equation}
and since the $(u_i,u_i^*)$ are a complete set of functions, we can
write
\begin{equation}
{\bar u}_j 
  = \sum_i \left(\alpha_{ji}u_i + \beta_{ji}u_i^*\right) .
\label{carBog3a}
\end{equation}
This relation is known as a Bogoliubov transformation, and 
the coefficients $\alpha_{ji}$ and $\beta_{ji}$ are Bogoliubov 
coefficients \cite{carBogoliubov}.  

We now have two vacuum states, one annihilated by the $a_i$ and one 
by the ${\bar a}_i$, and two number operators $N_i=a_i^\dagger a_i$ 
and ${\bar N}_i={\bar a}_i^\dagger{\bar a}_i$.  Using the 
orthonormality of the mode functions, it is straightforward to 
show that
\begin{equation}
\langle {\bar 0}| N_i |{\bar 0}\rangle = \sum_j |\beta_{ji}|^2 .
\label{carBog4}
\end{equation}
Thus if the coefficients $\beta_{ji}$ are not all zero, the 
``barred'' vacuum will have a nonvanishing ``unbarred'' particle 
content.

\index{Hawking radiation!from Bogoliubov transformation}%
In \cite{carHawking} and \cite{carHawkingb}, Hawking considered a
mass collapsing to form a black hole, and computed the Bogoliubov 
coefficients connecting an initial vacuum far outside the collapsing
matter to a final vacuum after the black hole formed.  He found 
that the ``barred'' observer at future null infinity will observe 
a thermal distribution of particles, with a temperature 
(\ref{carBHradiate3}).\footnote{The final distribution is actually
not quite thermal, but contains a ``greybody factor'' that reflects
the backscattering of some of the emitted radiation into the black 
hole.}  I will not go into details here; three very nice reviews can
be found in \cite{carJacobson,carTraschen,carVisser}.  The essential
physical feature is that ingoing vacuum modes ``pile up'' at the
horizon, giving an exponential relationship between ingoing and
outgoing surfaces of constant phase; the integrals that determine
the Bogoliubov coefficients $\beta_{ji}$ take the form
$$\int dv\,e^{i\omega v} e^{-i\frac{\omega}{\kappa}\ln v} ,$$
yielding gamma functions of complex arguments whose absolute squares
give the exponential behavior of a thermal distribution.

Hawking's derivation was based on a particular choice of vacuum 
state, but generalizations are possible.  For example, one may compare
the vacuum of a freely falling observer near the horizon to the
vacuum of an observer at future null infinity \cite{carUnruhb}.  One
can also look beyond the expectation value of the number operator, 
and express the full final state in terms of initial modes; one finds 
that it is exactly thermal \cite{carWaldb,carParker}.  Generalizations 
to spinor and gauge fields are straightforward, and yield the correct
fermionic and bosonic distribution functions.

\index{Unruh radiation}%
It is also possible to simplify the problem, by looking at the 
easier model of an accelerated observer in flat spacetime.  Such 
an observer is naturally described in Rindler coordinates 
\cite{carRindler}
$$ds^2 = e^{2a\xi}\left(d\eta^2 - d\xi^2\right) ,$$
in which the exponential relationship between the unaccelerated
and accelerated modes is easy to verify.  A straightforward
calculation of Bogoliubov coefficients shows that the accelerated
observer will see a thermal bath of ``Unruh radiation'' with a 
temperature $kT = \hbar a/2\pi$, where $a$ is proper acceleration 
\cite{carUnruhb}.  By the principle of equivalence, an observer
at rest near the horizon of a black hole should experience the
same effect, with the acceleration $a$ replaced by the appropriately
blue shifted surface gravity $\kappa$, the acceleration necessary
to hold the observer at rest.

\index{trans-Planckian modes}%
As I noted in the preceding section, the exponential relationship
between ``barred'' and ``unbarred'' modes may be a cause of concern.
The modes observed as Hawking radiation by an observer far from the 
black hole are red shifted from Planck-scale modes near the horizon, 
and it seems that one has extrapolated quantum field theory far beyond 
the range in which it is known to be valid.  To address this question,
a number of authors have looked at the effect of modifying the 
dispersion relations in a way that removes very high energy modes
(see, for example, \cite{carUnruh,carBrout,carCorley,carJacobsonc}).
For example \cite{carCorleyb}, one can replace the standard expression 
for the energy of a massless field, $\omega_{\bf k} = |{\bf k}|$, with
$$\omega_{\bf k}^2 = |{\bf k}|^2 - \frac{|{\bf k}|^4}{k_0{}^2},$$
eliminating modes with trans-Planckian energies.  Both numerical and 
analytical computations show that despite these drastic changes in the
high-frequency behavior, thermal Hawking radiation persists.  We now
have strong evidence that a few simple assumptions---a vacuum near
the horizon as seen in a freely falling frame, fluctuations that
start in the ground state, and adiabatic evolution of the modes---are
sufficient to guarantee thermal radiation \cite{carUnruhc}.

\subsubsection{Particle detectors in a black hole background}

The definitions of vacuum and particle number in the preceding 
section were taken from ordinary quantum field theory.  But finding 
observables in quantum gravity is notoriously difficult, and one 
might worry about the applicability of these definitions in a highly 
curved spacetime.  To address this issue, Unruh \cite{carUnruhb} 
and DeWitt \cite{carDeWitt} considered the response of a particle 
detector in a black hole background, and showed that such a 
detector sees thermal radiation at the Hawking temperature.  
Similarly, a static atom outside a black hole will be excited as 
one would expect in a thermal bath \cite{carYu}.

\subsubsection{The stress-energy tensor \label{carStress}}

\index{stress-energy tensor!black hole}%
One can obtain further invariant information about black hole 
radiation by evaluating the expectation value of the stress-energy 
tensor of a quantum field in a black hole background.  This is 
a large subject; good introductions can be found in the books 
\cite{carFrolov} and \cite{carBirrell}.  For these lectures, the 
most relevant result is that an ingoing negative energy flux at the 
horizon balances the outgoing flux of Hawking radiation observed 
at infinity, leading to a back-reaction in which the black hole's 
mass decreases (as expected from the intuitive argument of section 
\ref{carBHradiate}) and ensuring energy conservation.

The computation of $\langle T_{\mu\nu}\rangle$ in a black hole 
background is generally very difficult (see, for example, 
\cite{carPagec} or chapter 11 of \cite{carFrolov}). 
\index{trace anomaly}%
\index{conformal anomaly}%
\index{black hole!conformal description}%
In the special case of a massless scalar field---or more generally, 
a conformally invariant field---in two dimensions, the calculation 
drastically simplifies \cite{carChristensen}.  The key difference is 
that in two dimensions, conservation of the stress-energy tensor is 
sufficient to determine the full expectation value in terms of the 
trace anomaly $\langle T^\mu{}_\mu\rangle$, which, in turn, depends 
only on characteristics of the field in a flat background.  
\index{Hawking radiation!from conformal description}%
The resulting 
expectation values are thermal, and the total flux can be used to 
determine the temperature, which matches the Hawking temperature
(\ref{carBHradiate3}).

Quite recently, Robinson and Wilczek have shown how to extend this
result to more than two dimensions, by dimensionally reducing an
arbitrary field to two dimensions (or equivalently looking at a 
partial wave expansion) and trading the trace anomaly for a chiral 
anomaly \cite{carRobinson}.  Their method, with some variations 
(for example, \cite{carBanerjee}), has been quickly extended to a 
wide variety of black holes.  In a beautiful piece of work, Iso, 
Morita, and Umetsu have further shown that by looking at higher 
order correlators, one can use similar techniques to obtain not 
just the total flux, but the full blackbody spectrum of Hawking 
radiation \cite{carIso,carIsob}.

\subsubsection{Tunneling through the horizon \label{carTunnel}}

For many physical systems, we know that classically forbidden processes
can occur through quantum tunneling.  
\index{Hawking radiation!tunneling description}%
\index{tunneling}%
This is the case for Hawking 
radiation.  The idea of a tunneling description dates back to at
least 1975 \cite{carDamour}, but the nicest form is more recent, 
coming from Parikh and Wilczek's insight that one can think of the 
horizon tunneling past the emitted radiation rather than vice versa 
\cite{carParikh,carParikhb}.  

Consider a spherically symmetric system of mass $M$ consisting of 
a Schwarzschild black hole of mass $M-\omega$ emitting a shell of 
radiation of mass $\omega \ll M$.  In Painlev{\'e}-Gullstrand 
coordinates, chosen because they are stationary and nonsingular at 
the horizon, the shell moves in a spacetime with metric
$$ds^2 = \left(1-\frac{2G(M-\omega)}{r}\right)dt^2 
  - 2\sqrt{\frac{2G(M-\omega)}{r}}dt\,dr - dr^2 - r^2d\Omega^2 ,$$
and outgoing radial null geodesics satisfy
$${\dot r} = 1 - \sqrt{\frac{2G(M-\omega)}{r}} . $$
Now consider the imaginary part of the action for an outgoing positive
energy shell---to be interpreted as an s-wave particle---crossing 
the horizon from $r_{\scriptscriptstyle\mathit{in}}$ to 
$r_{\scriptscriptstyle\mathit{out}}$:
\begin{equation}
\mathop{Im} I = 
\mathop{Im}
\int_{r_{\scriptscriptstyle\mathit{in}}}^%
  {r_{\scriptscriptstyle\mathit{out}}}
p_r dr = 
\mathop{Im}
\int_{r_{\scriptscriptstyle\mathit{in}}}^%
  {r_{\scriptscriptstyle\mathit{out}}}
\int_0^{p_r}dp_r'\,dr =
\mathop{Im}\int_M^{M-\omega} 
\int_{r_{\scriptscriptstyle\mathit{in}}}^%
  {r_{\scriptscriptstyle\mathit{out}}}
\frac{dr}{{\dot r}}dH ,
\label{carTun1}
\end{equation}
where I have used Hamilton's equations of motion to write 
$dp_r = dH/{\dot r}$ and noted that the horizon moves inward from 
$GM$ to $G(M-\omega)$ as the particle is emitted.  Setting 
$H = M-\omega$ and inserting the value of $\dot r$ obtained from 
the null geodesic equation, one can perform the integral easily 
through a contour deformation, obtaining
\begin{equation}
\mathop{Im} I = 4\pi\omega G\left(M - \frac{\omega}{2}\right)
\label{carTun2}
\end{equation}
with 
$r_{\scriptscriptstyle\mathit{in}}>r_{\scriptscriptstyle\mathit{out}}$.
Again, the physical picture is that the horizon tunnels inward as the
black hole's mass decreases.

By standard quantum mechanics, the tunneling rate in the WKB 
approximation is then
\begin{equation}
\Gamma = e^{-2\mathop{Im} I/\hbar} 
 = e^{-8\pi\omega G\left(M - \frac{\omega}{2}\right)/\hbar} 
 = e^{\Delta S_{BH}}
\label{carTun3}
\end{equation}
where $\Delta S_{BH}$ is the change in the Bekenstein-Hawking entropy
(\ref{carintro1}).  By the first law of black hole mechanics, this is
$\hbar\omega/T_H$, and we recover thermal Hawking radiation.  

The tunneling derivation may be easily extended to other classes of 
black holes, and consistently reproduces the standard results.  Its 
relationship to Hawking's original derivation is not obvious, but 
Parikh and Wilczek note that the same analysis can describe a 
negative-energy particle tunneling into the black hole, thus offering 
a similar physical picture.

\subsubsection{Periodic Greens functions}

\index{thermal Greens functions}%
\index{Greens functions!thermal}%
Consider the two-point function of a scalar field $\varphi$ in a thermal
ensemble of inverse temperature $\beta$:
\begin{align}
G_\beta(x,0;x',t) 
 &= \mathop{Tr}\left(e^{-\beta H}\varphi(x,0)\varphi(x',t)\right)
 =\mathop{Tr}\left(\varphi(x,0)e^{-\beta H}e^{\beta H}\varphi(x',t)%
  e^{-\beta H}\right)\nonumber\\
 &=\mathop{Tr}\left(\varphi(x,0)e^{-\beta H}\varphi(x',t+i\beta)\right)
 = G_\beta(x',t+i\beta;x,0) ,
\label{carPerGreens1}
\end{align}
where I have used cyclicity of the trace and the fact that the 
Hamiltonian generates time translations, so 
$e^{\beta H}\varphi(x',t)e^{-\beta H} = \varphi(x',t+i\beta)$.  
In particular, (\ref{carPerGreens1}) implies that if a thermal Greens 
function is symmetric in its arguments, it must be periodic in time 
with period $i\beta$.  This argument may be run backwards, and such 
periodicity in imaginary time may be taken as the \emph{definition} 
of a thermal Greens function; in axiomatic quantum field theory, this 
is formalized as the KMS condition \cite{carKubo,carMartin,carHaag}.

\index{Unruh radiation}%
As early as 1976, Bisognano and Wichmann showed that the Greens 
function for a uniformly accelerated observer obeys the KMS condition 
\cite{carBis}.  By the equivalence principle, the same should hold 
for an observer at rest near the horizon of a black hole.  This is 
indeed the case, as shown by Gibbons and Perry \cite{carGibbonsb,%
carGibbonsc}, 
\index{Hawking temperature}%
who further demonstrated
that the periodicity corresponds exactly to the expected Hawking
temperature (\ref{carBHradiate3}).

\subsubsection{Gravitational instantons \label{carInstantons}}

The periodicity of Greens functions described above suggests that it 
might be worthwhile to consider the analytic continuation of black 
hole spacetimes to ``imaginary time.''  Near the horizon $r=r_+$, a 
stationary black hole metric takes the approximate form
$$ds^2 = 2\kappa(r-r_+)dt^2 - \frac{1}{2\kappa(r-r_+)}dr^2 
  - r_+^2d\Omega^2 .$$
Continuing to imaginary time $t=i\tau$ and replacing $r$ by the
proper distance
$$\rho = \frac{1}{\kappa}\sqrt{2\kappa(r-r_+)}$$
\index{black hole!Euclidean}%
\index{Euclidean black hole}%
to the horizon, we obtain the ``Euclidean black hole'' metric
\begin{equation}
ds^2 = d\rho^2 + \kappa^2\rho^2d\tau^2  + r_+^2d\Omega^2 .
\label{carGravinst1}
\end{equation}
The $\rho$--$\tau$ portion of this metric may be recognized as that 
of a flat two-plane in polar coordinates, with imaginary time $\tau$ 
serving as the angular coordinate.  The horizon $\rho=0$ has shrunk 
to a point.  To avoid a conical singularity at the origin, we must 
require that $\kappa\tau$ have period $2\pi$, i.e., that $\tau$ have 
period $2\pi/\kappa = 1/kT_{\scriptscriptstyle\mathit Hawking}$.

This result provides a simple way to understand the periodicity
of the Lorentzian Greens functions in imaginary time.  But it does 
more: it allows a steepest descent (``instanton'') approximation
to the gravitational path integral and a semiclassical derivation 
of the Bekenstein-Hawking entropy \cite{carGibbonsd}.  
\index{boundary terms}%
\index{Einstein-Hilbert action!boundary terms}%
The key ingredient
is the observation that on a manifold with boundary, the ordinary
Einstein-Hilbert action must be supplemented by a boundary term,
without which it may have no extrema \cite{carGibbonsd,carRegge}.  
At an extremum, the ``bulk'' contribution to the action,
$$\frac{1}{16\pi G}\int d^4x\,\sqrt{|g|}R ,$$
vanishes, but the boundary term can give a nonzero contribution.  
In the original work in this field, the boundary term was taken at 
infinity \cite{carGibbonsd,carHawkingc}, but it may more intuitively 
be placed at the origin of the Euclidean black hole, that is, at the 
horizon \cite{carBTZa,carTeitelboim,carHawkingHorowitz}.  This 
boundary term may be evaluated in a number of ways---a particularly 
elegant approach involves dimensional reduction to a disk in the 
$\rho$--$\tau$ plane \cite{carBTZa}---and yields an extremal action  
\begin{equation}
{\bar I}_{\scriptscriptstyle\mathit Euc} 
 = \frac{\ A_{\mathit\scriptstyle horizon}}{4\hbar G}
 - \beta(M + \Omega J + \Phi Q) .
\label{carGravinst2}
\end{equation}
\index{partition function}%
This Euclidean saddle point contributes 
$e^{{\bar I}_{\scriptscriptstyle\mathit Euc}}$ to the partition 
function, and from (\ref{carGravinst2}), we can recognize the result 
as the grand canonical partition function for a system with 
\index{Bekenstein-Hawking entropy}%
\index{entropy!black hole}%
entropy $S_{\scriptscriptstyle\mathit BH} 
= A_{\mathit\scriptstyle horizon}/4\hbar G$.  

These results can be extended to much more general stationary 
configurations containing horizons \cite{carHawkingHunter}.  The 
essential ingredient is a Killing vector with zeros, which become 
boundaries upon continuation to Euclidean signature.  One can also 
obtain an equivalent result by canonically quantizing the system 
while including the boundary terms; the boundary term at the 
horizon gives rise to a new term in the Wheeler-DeWitt equation, 
from which one can again recover the Bekenstein-Hawking entropy 
\cite{carCarTeit}.

\subsubsection{Black hole pair creation}

\index{pair creation!black holes}%
A further path integral derivation of black hole entropy comes 
from studying the spontaneous pair creation rate for black holes 
in a background magnetic field \cite{carGarfinkle}, electric 
field \cite{carBrown}, de Sitter space \cite{carMannRoss}, or more
complicated combinations of external fields \cite{carBooth}.  One
consistently finds that the production rate is enhanced by a factor
of $e^{S_{BH}}$, exactly the phase space factor one would expect for
a system in which the Bekenstein-Hawking entropy gives the logarithm
of the number of states.

\subsubsection{Quantum field theory and the eternal black hole
 \label{carEternal}}

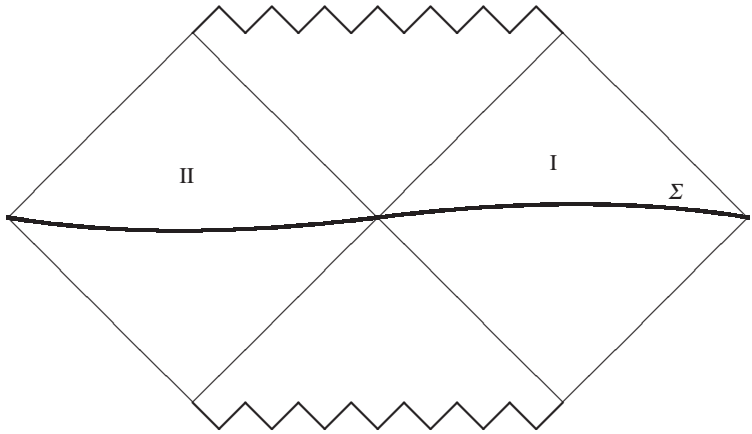
\begin{figure}
\begin{center}
\begin{picture}(200,160)(0,-5)
\put(10,10){\line(1,1){140}}
\put(150,10){\line(-1,1){140}}
\thicklines
\multiput(10,150)(20,0){7}{\line(1,1){10}}
\multiput(20,160)(20,0){7}{\line(1,-1){10}}
\multiput(10,10)(20,0){7}{\line(1,-1){10}}
\multiput(20,0)(20,0){7}{\line(1,1){10}}
\thinlines
\put(10,10){\line(-1,1){70}}
\put(-60,80){\line(1,1){70}}
\put(150,10){\line(1,1){70}}
\put(220,80){\line(-1,1){70}}
\linethickness{.4mm}
\qbezier(-60,80)(0,70)(80,80)
\qbezier(80,80)(160,90)(220,80)
\put(5,93){II}
\put(145,98){I}
\put(190,87){$\Sigma$}
\end{picture}
\caption{A Carter-Penrose diagram for an eternal black hole.
\label{carFig1}}
\end{center} 
\end{figure}

\index{density matrix!thermal}%
\index{thermal density matrix}%
Yet another derivation of Hawking radiation comes from considering
quantum field theory on an eternal black hole background.  Recall
that in Kruskal coordinates, a black hole spacetime splits into
four regions, as shown in figure \ref{carFig1}.
Consider a state defined on a Cauchy surface $\Sigma$ that passes
through the bifurcation sphere.  Region $II$ is invisible to an 
observer living in region $I$, so such an observer should trace over
the degrees of freedom in that region.  Even if the initial state 
is pure, such a trace will lead to a density matrix describing the
physics in region $I$.  This makes it plausible that the region $I$
observer will see thermal behavior, and detailed calculations show
that this is indeed the case.

In particular, for a free quantum field there is at most one quantum 
state, the Hartle-Hawking vacuum state, that is regular everywhere 
on the horizon \cite{carWaldbk,carKay}.  
\index{Hawking temperature}%
For a scalar field, a direct computation 
shows that the density matrix obtained by tracing over region $II$ 
is thermal, with a temperature $T_{\scriptscriptstyle\mathit Hawking}$ 
\cite{carIsraelb}.  For more general fields, the same can be shown 
by means of fairly sophisticated quantum field theory 
\cite{carWaldbk,carKay}, or by general path integral arguments 
\cite{carJacobsond}.

\subsubsection{Quantum gravity in 2+1 dimensions \label{car3D}}

Most standard derivations of black hole thermodynamics hold in an
arbitrary number of dimensions, with changes only in the greybody 
factors for Hawking radiation.  In three spacetime dimensions,
though, many approaches become much simpler.  
\index{BTZ black hole}%
\index{black hole!BTZ}%
\index{black hole!2+1 dimensions|see{black hole, BTZ}}%
The BTZ solution
\cite{carBTZ,carCarlipa} is a vacuum solution of the Einstein field 
equations in 2+1 dimensions with a negative cosmological constant.  
It has all of the standard features of a rotating black hole---an 
event horizon, an inner Cauchy horizon, the same causal structure 
as that of a (3+1)-dimensional asymptotically anti-de Sitter black 
hole---but is, at the same time, a space of constant negative 
curvature.  This latter feature greatly simplifies many 
derivations: for example, Greens functions can be computed 
exactly and their periodicity in imaginary time exhibited 
explicitly (see \cite{carCarlipa} for a review).  As was first 
suggested in \cite{carCarlipb}, it might even be possible to use
the relationship between three-dimensional general relativity and
two-dimensional conformal field theory \cite{carWitten} to find an
exact description of the quantum states of the BTZ black hole; the
present status of this conjecture is discussed in \cite{carCarlipc}.

The simplicity of the (2+1)-dimensional setting also permits 
an approach that is not readily available in higher dimensions.  
The methods I have described so far are based on properties of 
quantum fields in a classical, or at best semiclassical, black 
hole background.  In three dimensions, one can work in the opposite 
direction, starting with a \emph{quantum} black hole coupled to 
a \emph{classical} source.  As I shall discuss further in section 
\ref{carStates}, three-dimensional gravity with a negative 
cosmological constant is closely related to a two-dimensional 
field theory living at the ``boundary'' of asymptotically anti-de 
Sitter space. 
\index{Hawking radiation!from detailed balance}%
Emparan and 
Sachs have shown how to couple this two-dimensional field theory 
to a classical scalar field, allowing the computation of transition 
rates among black hole states due to emission and absorption of 
the classical field \cite{carEmparan}.  By using detailed balance 
arguments, they recover standard Hawking radiation, including 
the correct greybody factors, from this fundamentally quantum 
gravitational picture.

\subsubsection{Other microscopic approaches}

The derivations I have described so far are essentially 
``thermodynamic,'' based on macroscopic properties of black holes.  
As I shall discuss in the following sections, we now also have a 
large number of ``statistical mechanical'' derivations, based on 
analyses of the microscopic states of the black hole.  These 
microscopic approaches are not complete---string theory derivations, 
for example, are most reliable for extremal and near-extremal black 
holes, while loop quantum gravity derivations contain an order one 
parameter that, so far, must be adjusted by hand---but they seem 
to work well within their ranges of validity.  When combined with 
the macroscopic approaches above, they provide strong evidence for 
the reality of black hole thermodynamics.

\section{Black Hole Statistical Mechanics}

In ordinary thermodynamic systems, thermal properties are macroscopic 
reflections of the underlying microscopic physics.  Temperature is a
measure of the average energy of the constituents of a system, for 
instance, while entropy is essentially the logarithm of the number 
of states with specified macroscopic properties.  The connection 
between the microscopic and macroscopic properties, given by 
statistical mechanics, has been remarkably successful across physics.

\index{black hole!microstates}%
Given the thermodynamic properties of black holes, it is natural to
ask whether these, too, have a statistical mechanical interpretation.
Such an explanation would almost certainly involve quantum 
gravity---the Bekenstein-Hawking entropy (\ref{carintro1}) involves
both Planck's constant $\hbar$ and Newton's constant $G$---and we 
might hope to learn something about the deep mysteries of quantum 
gravity.

To find such a statistical mechanical description, one should, in 
principle, carry out a number of steps:
\begin{enumerate}
\item find a candidate quantum theory of gravity (not an easy task);
\item identify black holes in the theory (also not easy);
\item identify observables such as horizon area (surprisingly hard%
 ---finding physical observables in a quantum theory of gravity is 
 notoriously difficult \cite{carCarliprev});
\item count the microstates for a black hole configuration (perhaps 
 easier, but still not trivial); 
\item compare to the Bekenstein-Hawking entropy (perhaps relatively
 easy);
\item compute interactions with external fields, evaluate Hawking
 radiation, etc.\ (not at all easy); 
\item try to identify new quantum gravitational effects (the horizon
 area spectrum? evaporation remnants? higher order corrections to 
 the Bekenstein-Hawking entropy? correlations across the horizon?).
\end{enumerate}

Until recently, these steps seemed far beyond reach.  In 1996,
though, Strominger and Vafa published a remarkable paper in which 
they explicitly computed the entropy of a class of extremal black
holes in string theory from the microscopic quantum theory 
\cite{carStrominger}.  Since then, a flood of new microscopic 
derivations of black hole thermodynamics has appeared.  The new 
puzzle---the ``problem of universality''---is that although these
derivations seem to be using very different methods to count very 
different states, they all obtain the same thermodynamic properties.

\subsection{The many faces of black hole statistical mechanics}

In this section, I will briefly review some of the statistical 
mechanical approaches to black hole thermodynamics, and in 
particular the Bekenstein-Hawking entropy.  As in section 
\ref{carManyT}, I will not go into detail, but will instead 
try to provide an overall flavor of the work, along with 
references for further study.

\subsubsection{String theory: weakly coupled strings and branes 
\label{carWeak}}

\index{string theory!weakly coupled}%
\index{string theory!black hole entropy}%
\index{black hole entropy!string theory}%
The first breakthrough in the counting of black hole microstates 
came with the work of Strominger and Vafa on extremal black holes 
in string theory \cite{carStrominger}.  Their approach can be 
summarized as follows.  

The effective low-energy field theory coming from string theory 
contains a number of gauge fields, each of which can give a charge 
to a black hole.  An extremal supersymmetric (BPS) black hole is 
uniquely characterized by its charges; in particular, its horizon 
area can be expressed in terms of these charges.  Given such a 
black hole, one can imaging tuning down the couplings, weakening
gravity until the black hole ``dissolves'' into a gas of weakly 
coupled strings and branes.  In this weakly coupled system, the 
charges can be expressed in terms of the number of strings and 
branes and the quantized momentum carried by strings.  Furthermore, 
the states---the excitations of the string-brane system---can be 
explicitly counted \cite{carMathur0}.  We can therefore write the 
number of states in terms of the numbers of strings and branes, 
and thus the charges.  
\index{Bekenstein-Hawking entropy}%
\index{entropy!black hole}%
Comparing this number to the 
horizon area, we recover the standard Bekenstein-Hawking entropy as 
the logarithm of the number of states.

One might worry that the number of states might not be the same in 
the weakly coupled system as in the strongly coupled black hole.  
For the supersymmetric case, though, this number is protected by 
nonrenormalization theorems.  For black holes far from extremality, 
on the other hand, the computations are much more difficult; there 
are qualitative arguments that give an entropy proportional to the 
horizon area, but the exact proportionality factor of $1/4$ is 
difficult to obtain \cite{carSusskind,carHorowitzPol}.

It was quickly realized that the Strominger-Vafa results could be 
extended to a wide variety of extremal and near-extremal black 
holes, and through duality relations to a number of nonextremal 
black holes as well.   Nice reviews can be found in \cite{carPeet} 
and \cite{carDas}; for recent progress on the four-dimensional 
Kerr black hole, see \cite{carHorowitz}.

This string theory approach has been remarkably successful, 
determining not only the Bekenstein-Hawking entropy for extremal 
and near-extremal black holes, but also describing their 
interactions with other fields and their emission of Hawking 
radiation.  
\index{universality!problem of}%
The method has one peculiarity,
though, to which I will return below.  Suppose you ask me for the 
entropy of a three-charge black hole in five dimensions.  I will 
compute the horizon area in the strongly coupled theory in terms 
of the charges, compute the number of states in the weakly coupled 
theory in terms of the charges, compare the two, and reply that 
the entropy is one-fourth of the horizon area.  If you now ask 
me for the entropy of a four-charge black hole, or a black hole 
in six dimension, I cannot simply tell you that it is one-fourth 
of the horizon area; I must recompute the horizon area and the 
number of states in terms of the new parameters and compare the 
answers again.  Each new black hole requires a new calculation:
the theory tells us that the number of microstates of a black hole 
matches the Bekenstein-Hawking entropy (\ref{carintro1}), but it 
tells us so one black hole at a time.

\subsubsection{String theory: ``fuzzballs''}

\index{black hole entropy!fuzzballs}%
One can run the argument in the preceding section backwards:
given a particular excitation of the weakly coupled string and 
brane system, one can ask exactly what geometry results at strong 
coupling.  The result is a ``fuzzball'' picture, in which particular 
black hole states correspond to complicated geometries that have 
\emph{no} horizon or singularity, but that look very much like black 
hole geometries outside the would-be horizon \cite{carMathur,%
carMathurb}.  In special cases, one can count the number of such 
``fuzzball'' geometries and reproduce the Bekenstein-Hawking entropy, 
and it seems likely that this result can be extended to more general 
black holes, although it is an open question whether simple geometric 
descriptions will always suffice \cite{carSkenderis}.  Samir Mathur 
has discussed this approach extensively in his lectures, to which I 
refer the reader \cite{carMathur0}.

\subsubsection{String theory: the AdS/CFT correspondence 
 \label{carAdSCFT}}

\index{AdS/CFT correspondence}%
\index{black hole entropy!AdS/CFT correspondence}%
Yet another string theory approach to black hole statistical
mechanics is based on Maldacena's celebrated AdS/CFT correspondence 
\cite{carMalda,carAGMOO}.  This very well-supported conjecture 
asserts a duality between string theory in $d$-dimensional 
asymptotically anti-de Sitter spacetime and a conformal field 
theory in a flat $(d-1)$-dimensional space that can, in a sense, 
be viewed as the boundary of the AdS spacetime.  
\index{holographic}%
This correspondence is naturally ``holographic'' (see section 
\ref{carHolography}), describing the black hole in terms of a 
lower-dimensional theory and thus offering a framework for
understanding the dependence of entropy on area rather than volume.

\index{BTZ black hole}%
\index{black hole!BTZ}%
For asymptotically anti-de Sitter black holes, this correspondence 
makes it possible to compute entropy by counting states in a 
(nongravitational) dual conformal field theory.  The simplest 
case is the (2+1)-dimensional BTZ black hole discussed in section 
\ref{car3D}, whose dual is a two-dimensional conformal field theory.  
\index{conformal anomaly}%
\index{black hole!conformal description}%
As I shall 
discuss in section \ref{carCardyformula}, the density of states in 
such a theory has an asymptotic behavior controlled by a single 
parameter, the central charge $c$.  For asymptotically anti-de Sitter 
gravity in 2+1 dimensions, this central charge is dominated by a 
classical contribution, which was discovered some time ago by Brown 
and Henneaux \cite{carBrownHen}.  Strominger \cite{carStromingerb} 
and Birmingham, Sachs, and Sen \cite{carBSS} independently realized 
that this result could be used to compute the BTZ black hole entropy, 
\index{Bekenstein-Hawking entropy}%
\index{entropy!black hole}%
reproducing the Bekenstein-Hawking expression.

While this result applies directly only to the special case of 
three-dimensional spacetime, it has an important generalization.  
Many of the higher dimensional near-extremal black holes of string 
theory---including black holes that are not themselves asymptotically 
anti-de Sitter---have a near-horizon geometry of the general form 
$\mathit{BTZ}\times\mathit{trivial}$,
where the ``trivial'' part merely renormalizes constants in the
calculation of entropy.  As a consequence, the BTZ results can be 
used to find the entropy of a large class of stringy black holes, 
including most of the black holes whose states can be counted 
in the weak coupling approach of section \ref{carWeak} 
\cite{carSkenderisb}.

\subsubsection{Loop quantum gravity \label{carLoop}}

\index{spin networks}%
In the quest for quantum gravity, the leading alternative to string 
theory is loop quantum gravity \cite{carRovelli}.  The fundamental 
``position'' variable in this theory is a three-dimensional 
$\mathrm{SU}(2)$ connection; a state is a complex-valued function 
of (generalized) connections.  A useful basis of states consists 
of spin networks, graphs with edges labeled by $\mathrm{SU}(2)$ 
representations (``spins'') and vertices labeled by intertwiners.  
A spin network state can be evaluated on a given connection to give 
a complex number by computing the holonomies along the edges in the
specified representations and combining them with the intertwiners
at the vertices.

Given a surface $\Sigma$, one can define an area operator 
${\hat A}_\Sigma$ that acts on loop quantum gravity states.  It 
may be shown that spin networks are eigenfunctions of these 
operators, with eigenvalues of the form
$$A_\Sigma = 8\pi\gamma G \sum_j \sqrt{j(j+1)} ,$$
where the sum is over the spins $j$ of edges of the spin network 
that cross $\Sigma$.  
\index{Barbero-Immirzi parameter}%
\index{Immirzi parameter}%
The parameter $\gamma$, the Barbero-Immirzi parameter, 
represents a quantization ambiguity, and its physical significance 
is poorly understood; theories with different values of $\gamma$ 
may be inequivalent, but it has been suggested that $\gamma$ may 
not appear in properly renormalized observables \cite{carJacobsone} 
or in a slightly different approach to quantization 
\cite{carAlexandrov}.

\index{black hole entropy!loop quantum gravity}%
Given this structure, a natural first attempt to count black hole 
states is to enumerate inequivalent spin networks crossing the 
horizon that yield a specified area \cite{carKrasnov,carRovellib}. 
The more careful variation of this idea \cite{carABCK,carABK} takes 
into account the fact that when one restricts to a black hole 
spacetime, one must place ``boundary conditions'' on the horizon 
to ensure that it is, in fact, a horizon.  These conditions, in 
turn, require the addition of boundary terms to the Einstein-Hilbert
action, which induce a three-dimensional Chern-Simons action on 
the horizon.  The number of states of this Chern-Simons theory is 
closely related to the number of spin networks that induce the 
correct horizon area, but with slightly more subtle combinatorics.  
The ultimate result is that the black hole entropy takes the form 
\cite{carDomagala,carMeissner}
\begin{equation}
S = \frac{\gamma_M}{\gamma}
  \frac{\ A_{\mathit\scriptstyle horizon}}{4\hbar G} ,
\label{carLoopQG1}
\end{equation}
where $\gamma$ is the Barbero-Immirzi parameter and
$$\gamma_M \approx .23753$$
is a numerical constant determined as the solution of a particular
combinatoric problem.  
\index{Bekenstein-Hawking entropy}%
\index{entropy!black hole}%
If one chooses $\gamma=\gamma_M$, one thus recovers
the standard Bekenstein-Hawking entropy.  

The physical significance of this rather peculiar value of the 
Barbero-Immirzi parameter is not understood, and it may reflect an 
inadequacy in the quantization procedure or the definition of the 
area operator \cite{carAlexandrov}.  Note, though, that $\gamma$ 
only appears in the combination $G\gamma$, so this choice may be 
viewed as a finite renormalization of Newton's constant.  If the 
same shift occurs in the attraction between two masses, its 
interpretation becomes straightforward.  Unfortunately, the 
Newtonian limit of loop quantum gravity is not yet well enough 
understood to see whether this is the case.

In any case, though, once $\gamma$ is fixed for one type of black
hole---the static Schwarzschild solution, say---the loop quantum 
gravity computations give the correct entropy for a wide variety 
of others, including charged black holes, rotating black holes, 
black holes with dilaton couplings, black holes with higher genus 
horizons, and black holes with arbitrarily distorted horizons 
\cite{carAshtekarLew,carEngle}.  In particular, there is no need 
to restrict oneself to near-extremal black holes.  Hawking radiation, 
on the other hand, is not yet very well understood in this approach, 
although there has been some progress \cite{carBarreira,carKrasnovb}.

An alternate approach to black hole entropy also exists within 
the framework of loop quantum gravity \cite{carLivine}.  Here, one 
again looks at a horizon area determined by edges of a spin network, 
but instead of counting states in an induced boundary theory, one
merely asks the number of ways the spins can be joined to a single 
interior vertex.  This amounts, in essence, to completely 
coarse-graining the interior state of the black hole, and is 
comparable in spirit to the thermodynamic derivation of section 
\ref{carEternal}.  One again obtains an entropy proportional to 
the horizon area, although with a different value of the 
Barbero-Immirzi parameter.

\subsubsection{Induced gravity}

In 1967, Sakharov suggested that the Einstein-Hilbert action for
gravity might not be fundamental \cite{carSakharov}.  If one 
starts with a theory of fields propagating in a curved spacetime, 
counterterms from renormalization will automatically induce a 
gravitational action, which will almost always include an 
Einstein-Hilbert term at lowest order \cite{carAdler}.  
Gravitational dynamics would then be, in Sakharov's terms, a 
sort of ``metric elasticity'' induced by quantum fluctuations.

\index{black hole entropy!induced quantum gravity}%
One can write down an explicit set of ``heavy'' fields that 
can be integrated out in the path integral to induce the 
Einstein-Hilbert action.  By including nonminimally coupled 
scalar fields, one can obtain finite values of Newton's constant 
and the cosmological constant.  It is then possible to go back
and count states of the heavy fields in a black hole background 
\cite{carFrolovb}.
\index{Bekenstein-Hawking entropy}%
\index{entropy!black hole}%
The nonminimal couplings lead to some subtleties in the definition
of entropy, but in the end the computation reproduces the standard
Bekenstein-Hawking value.  Furthermore, the reduction to a 
two-dimensional conformally invariant system near the horizon, in
the spirit of the thermodynamic approach of section \ref{carStress},
allows a counting of states by standard methods of conformal field
theory \cite{carFrolovc}.  We thus obtain a new, and apparently 
quite different, view of the microstates of a black hole as those 
of the ordinary quantum fields responsible for inducing the
gravitational action.

\subsubsection{Entanglement entropy}

\index{density matrix!thermal}%
\index{thermal density matrix}%
\index{entanglement entropy}%
\index{black hole entropy!entanglement}%
As discussed in section \ref{carEternal}, one way to obtain the
thermodynamic properties of a black hole is to trace out the
degrees of freedom behind the horizon, generating a density matrix
for the external observer from a globally pure state.  This process 
also produces a quantum mechanical ``entanglement entropy,'' which 
can be thought of as a measure of the loss of information about 
correlations across the horizon.  The suggestion that this 
entanglement entropy might account for the Bekenstein-Hawking 
entropy is an old one \cite{carBombelli,carSrednicki}, and it is 
not hard to show that for many (although not all \cite{carRequardt}) 
states, the entanglement entropy is proportional to the horizon 
area: the main contribution comes from correlations among degrees 
of freedom very close to the horizon, and does not involve ``bulk'' 
degrees of freedom.  The \emph{coefficient} of this entropy, on 
the other hand, is infinite, and must be cut off \cite{cartHooft}, 
leading to an expression that depends strongly on both the 
nongravitational content of the theory (the number and species 
of ``entangled'' fields contributing to the entropy) and the value 
of the cutoff.  

The same modes that cause the entanglement entropy to diverge also 
give divergent contributions to the renormalization of Newton's 
constant, and it has been suggested that the two divergences may 
compensate \cite{carSusskindb}.  
\index{holographic}%
\index{AdS/CFT}%
This notion has recently gained 
new life with a proposal by Ryu and Takayanagi for a ``holographic'' 
description of entanglement entropy \cite{carRyu,carHubeny}, in 
which the $d$-dimensional spacetime containing a black hole is 
embedded at the asymptotic boundary of $(d+1)$-dimensional anti-de 
Sitter space.  The idea is inspired by the string theory AdS/CFT 
correspondence, and can be largely proven to work in situations 
in which such a correspondence exists \cite{carFursaev}; the bulk 
anti-de Sitter metric provides a natural cutoff, yielding finite 
contributions to both $S$ and $G$.  When applied to a black hole, 
\index{Bekenstein-Hawking entropy}%
\index{entropy!black hole}%
the proposal correctly reproduces the 
standard Bekenstein-Hawking entropy \cite{carEmparanb}, providing 
yet another physical picture of the relevant microstates.

\subsubsection{Other approaches}

A variety of other microscopic descriptions of black hole 
thermodynamics have also been proposed.  In the causal set 
formulation of quantum gravity---in which a continuous spacetime 
is replaced by a discrete set of points with prescribed causal 
relations---there is evidence that the Bekenstein-Hawking entropy 
is given by the number of points in the future domain of dependence 
of a spatial cross-section of the horizon \cite{carRideout}.  York 
has estimated the entropy obtained by quantizing the quasinormal 
modes \cite{carSiopsis} of the Schwarzschild black hole, finding 
a result that lies within a few percent of the Bekenstein-Hawking 
value \cite{carYork}.  Black hole entropy can be related to the 
Kolmogorov-Sinai entropy of a string spreading out on the black 
hole horizon \cite{carRopotenko}.  A number of mini- and 
midisuperspace models---models in which most of the degrees of 
freedom of the gravitational field are frozen out---have also been 
proposed to explain black hole statistical mechanics 
\cite{carVaz,carMakela,carKiefer}, though none is yet very 
convincing.

One can also build ``phenomenological'' models of black hole 
microstates, in which the horizon area is simply assumed to be 
quantized \cite{carBekensteind,carKastrup,carBarvinsky,carBekensteine}.  
Such models do not, of course, tell us \emph{why} area is quantized, 
and thus do not address the fundamental physical questions of black 
hole statistical mechanics, but they can suggest useful directions 
for further research.  
 
\index{quasinormal modes!area spectrum and}%
Suppose, for example, that the black hole area spectrum is discrete 
and equally spaced, and that the exponential of the entropy 
(\ref{carintro1}) gives an exact count of the number of states 
at a given horizon area.  Then the difference between two adjacent 
values must be an integer; that is,
\begin{equation}
\Delta A = 4\hbar G \ln k
\label{carOther1}
\end{equation}
for some integer $k$.  Hod has pointed out \cite{carHod} that for 
the Schwarzschild black hole, the most highly damped quasinormal 
modes \cite{carSiopsis}---the damped ``ringing modes'' of an excited 
black hole---have frequencies whose real part approaches
$$\mathop{Re}\omega = \ln3/8\pi GM$$
(a numerical result later verified analytically \cite{carMotl}).  If
one applies the Bohr correspondence principle and argues that area
eigenstates of the black hole should change by emission of quanta of
energy $\hbar\omega$, one obtains
$$\Delta A = 32\pi G^2M\Delta M = 4\hbar G\ln 3 ,$$
matching (\ref{carOther1}) with $k=3$.  It is not yet clear whether 
this result has deep significance.  It seems to extend to general 
single-horizon black holes \cite{carDaghigh} and in a more complicated 
way to many ``stringy'' black holes \cite{carBirminghamCar}, but 
results for charged and rotating black holes are unclear (for an 
optimistic view, see \cite{carHodb}).

\index{Bekenstein-Hawking entropy}%
\index{entropy!black hole}%
One can also describe the Bekenstein-Hawking entropy as a count 
of the number of distinct ways that a black hole with specified 
macroscopic properties can be made from collapsing matter 
\cite{carZurek}.  Like the phenomenological models of area 
quantization, this result does not really describe the microscopic 
degrees of freedom of the black hole itself (except perhaps in the 
``membrane paradigm'' \cite{carThorne}), but it strongly suggests 
that if the formation of a black hole is a unitary process, such 
degrees of freedom must exist.

\section{The problem of universality}

\index{universality!problem of}%
One of the main lessons of the preceding section is that a great 
many different models of black hole microphysics yield the same 
thermodynamic properties.  Some of these models are clearly ad hoc, 
but others are carefully worked out consequences of serious approaches 
to quantum gravity.  So the new question is why everyone is getting 
the same answer.

To some extent, this ``problem of universality'' is a selection 
issue: there are undoubtedly computations that gave the ``wrong'' 
answer for black hole entropy and were discarded without being 
published.  But as noted in section \ref{carWeak}, even within 
a particular well-motivated and successful string theory model 
we do not yet understand the universality of the entropy-area 
relationship.  And regardless of what one may think about any 
one particular approach, one must still explain why \emph{any} 
microscopic model reproduces the results of Hawking's original 
thermodynamic computation, a computation that seems to require no 
information about quantum gravity at all.

There are other situations, of course, in which thermodynamic 
properties do not depend too delicately on an underlying quantum 
theory.  For example, for a large range of parameters the entropy 
of a box of gas depends only very weakly on whether the molecules 
are fermions or bosons.  But in cases like this, we have a 
\emph{classical} microscopic description, and the correspondence 
principle guarantees that the quantum theory will give a good 
approximation for the classical results.  For a black hole, things 
are different: the only classical description we have is one in
which black holes have no hair---no phase space volume---and thus 
no entropy.   We need something new, some new principle that 
determines the quantum mechanical density of states in terms of 
the classical characteristics of a black hole.

I do not know the ultimate explanation for this universal behavior, 
but in the remainder of this section, I will make a tentative 
suggestion and offer some evidence that it may be correct.

\subsection{The Cardy formula \label{carCardyformula}}

\index{black hole!conformal description}%
\index{conformal anomaly}%
\index{Cardy formula}%
I only know of one well-understood case in which universality of 
the sort we see in black hole statistical mechanics appears 
elsewhere in physics.  Consider a two-dimensional conformal field 
theory, that is, a theory in two spacetime dimensions that is 
invariant under diffeomorphisms (``generally covariant'') and Weyl 
transformations (``locally scale invariant'').  If we choose complex 
coordinates $z$ and ${\bar z}$, the basic symmetries of such a 
theory are the holomorphic and antiholomorphic diffeomorphisms 
$z\rightarrow f(z)$, ${\bar z}\rightarrow{\bar f}({\bar z})$.  
These are canonically generated by 
\index{Virasoro algebra}%
``Virasoro generators'' $L[\xi]$ and ${\bar L}[{\bar \xi}]$ 
\cite{carCFT}.  Such a theory has two conserved charges, 
$L_0 = L[\xi_0]$ and ${\bar L}_0 = {\bar L}[{\bar\xi}_0]$, 
which can be thought of as ``energies'' with respect to constant 
holomorphic and antiholomorphic transformations, or alternatively
as linear combinations of energy and angular momentum.

As generators of diffeomorphisms, the Virasoro generators have an 
algebra that is almost unique \cite{carTeitelboimb}: 
\begin{align}
&\left\{L[\xi],L[\eta]\right\} = L[\eta\xi' - \xi\eta']
 + \frac{c}{48\pi}\int dz\left( \eta'\xi'' - \xi'\eta''\right)
 \nonumber \\
&\left\{L[\xi],{\bar L}[{\bar\eta}]\right\} = 0
 \label{carCardyform1} \\
&\left\{{\bar L}[{\bar\xi}],{\bar L}[{\bar\eta}]\right\} 
 = {\bar L}[{\bar\eta}{\bar\xi'} - {\bar\xi}{\bar\eta'}]
 + \frac{{\bar c}}{48\pi}\int d{\bar z}\left({\bar\eta}'{\bar\xi}''
 - {\bar\xi}'{\bar\eta}''\right)\nonumber .
\end{align}
The central charges $c$ and $\bar c$ determine the unique central 
extension of the ordinary algebra of diffeomorphisms.  These 
constants can occur classically, coming, for instance, from 
boundary terms in the generators \cite{carBrownHen}, or can appear 
upon quantization.

Now consider a conformal field theory for which the lowest 
eigenvalues of $L_0$ and ${\bar L}_0$ are nonnegative numbers 
$\Delta_0$ and ${\bar\Delta}_0$.  In 1986, Cardy discovered a 
remarkable result \cite{carCardy,carCardyb}: the density of states 
$\rho(\Delta,\bar\Delta)$ at eigenvalues $(\Delta,{\bar\Delta})$ 
of $L_0$ and ${\bar L}_0$ has the simple asymptotic behavior
\begin{equation}
\ln\rho(\Delta,\bar\Delta) \sim 2\pi\left\{
 \sqrt{\frac{c_{\hbox{\scriptsize\it eff}}\Delta}{6}} 
 + \sqrt{\frac{{\bar c}_{\hbox{\scriptsize\it eff}}{\bar\Delta}}{6}}\,
 \right\}, \ \ \ \hbox{with}\ \ 
 c_{\hbox{\scriptsize\it eff}} = c-24\Delta_0, \ 
 {\bar c}_{\hbox{\scriptsize\it eff}} = {\bar c}-24{\bar\Delta}_0 .
\label{carCardyform2}
\end{equation}
The entropy is thus determined by the symmetry, independent of any 
other details---exactly the sort of universality we are looking for.

\index{black hole!conformal description}%
A typical black hole is neither two-dimensional nor conformally 
invariant, of course, so this result may at first seem irrelevant.  
But there is a sense in which black holes become \emph{approximately} 
two-dimensional and conformal near the horizon.  For fields in a 
black hole background, for instance, excitations in the $r$--$t$ 
plane become so blue shifted relative to transverse excitations and 
dimensionful quantities that an effective two-dimensional conformal 
description becomes possible \cite{carBirm,carGupta,carCamblong}.  
Indeed, as noted in section \ref{carStress}, the full Hawking 
radiation spectrum can be derived from such an effective description 
\cite{carIso,carIsob}.  Martin, Medved, and Visser have further 
shown that a generic near-horizon region has a conformal symmetry, 
in the form of an approximate conformal Killing vector 
\cite{carMartinMed,carMartinMedb}.

\subsection{Horizons and constraints}

\index{BTZ black hole}%
\index{black hole!BTZ}%
For the special case of the (2+1)-dimensional BTZ black hole, 
the Cardy formula can be used directly to count states.  For 
this solution, the boundary at infinity is geometrically a 
two-dimensional flat cylinder, and the asymptotic diffeomorphisms 
that respect boundary conditions satisfy a Virasoro algebra with 
a classical central charge \cite{carBrownHen}, which can be used 
in the Cardy formula \cite{carStromingerb,carBSS}.  As described 
in section \ref{carAdSCFT}, this calculation can be extended to 
a number of near-extremal black holes whose near-horizon geometry 
contains a $\mathit{BTZ}$ factor.  For more general black holes, 
though, something new is needed.

\index{horizon constraints}%
One key question, I believe, is how to specify that one is talking 
about a black hole in quantum gravity.  One cannot simply require 
a fixed metric: the components of the metric do not all commute, 
and cannot be simultaneously specified in a quantum theory.  For
the BTZ case, the key element is a set of boundary conditions at
infinity, but in general it seems more natural to consider conditions
at the horizon.  Two 
approaches to this question are currently under investigation, 
each leading to an effective two-dimensional conformal description 
in which the Cardy formula might be applicable.

\subsubsection{The horizon as a boundary}

\index{Einstein-Hilbert action!boundary terms}%
\index{boundary terms}%
\index{black hole!conformal description}%
\index{Cardy formula}%
The first approach \cite{carCarlipd,carCarlipe} is to introduce 
``boundary conditions'' at the horizon.  The horizon is not, of 
course, a genuine boundary, but it is a place at which we must 
restrict the value of the metric, precisely to ensure that it is a 
horizon.  As in the BTZ case, such a restriction forces us to add 
new boundary terms to the canonical generators of diffeomorphisms, 
changing their algebra.  One finds a conformal symmetry in the 
$r$--$t$ plane with a classical central charge.  For a large
variety of black holes, 
\index{Bekenstein-Hawking entropy}%
\index{entropy!black hole}%
it has been shown that the Cardy formula  
then yields the correct entropy.\footnote{For this section, see 
\cite{carCarlipf} for further references.}  

On the other hand, the diffeomorphisms whose algebra yields that 
central charge, essentially those that leave the lapse function 
invariant, are generated by vector fields that blow up at the 
horizon.  This is not necessarily a bad thing---from the perspective
of an external observer, many physical quantities diverge at the
horizon---but the status of these transformations is not clear.  In
addition, the ``horizon as boundary'' method has trouble with the 
two-dimensional black hole, and some normalization issues are not
completely sorted out.  
\index{Virasoro algebra}%
A related approach is to look for approximate conformal symmetry 
near the horizon \cite{carSolo,carCarlipg}; one again finds a 
Virasoro algebra with a central charge that seems to lead to the 
correct entropy, but there are again some normalization ambiguities.

\subsubsection{Horizon constraints \label{carHorizoncon}}

\index{black hole!conformal description}%
\index{Cardy formula}%
\index{horizon constraints}%
A more recent approach \cite{carCarliph,carCarlipi} is to impose 
the presence of a horizon by adding ``horizon constraints'' in 
the canonical formulation of gravity, that is, introducing new 
constraints that restrict data on a specified surface to be that 
of a black hole horizon.  In outline, the procedure is this:
\begin{enumerate}
\item dimensionally reduce to the two-dimensional $r$--$t$ plane near 
the horizon;
\item continue to Euclidean signature, shrinking the horizon to a point
as in section \ref{carInstantons}, and evolve radially;
\item impose constraints on a small circle around the horizon that
force the initial data be that of a ``stretched horizon'';
\item adjust the diffeomorphism constraints on the stretched horizon
a la Bergmann and Komar \cite{carBergmann,carDirac,carDiracb} to 
make them commute with the new horizon constraints;
\item find the resulting algebra and central charge.
\end{enumerate}
\index{Bekenstein-Hawking entropy}%
\index{entropy!black hole}%
The Cardy formula again reproduces the correct Bekenstein-Hawking 
entropy.

\subsubsection{Universality again}

\index{universality!problem of}%
If either of these approaches is to be an answer to the ``problem 
of universality,'' it must be that the horizon conformal 
symmetries are secretly present in the various other computations 
of black hole entropy.  I do not know whether this is the case; it 
is a subject of  continuing research.  

One fairly simple test is to compare the near-horizon Virasoro 
algebra of section \ref{carHorizoncon} with the asymptotic 
Virasoro algebra of the BTZ black hole, which is the key element 
in the AdS/CFT computations of section \ref{carAdSCFT}.  It is 
shown in \cite{carCarlipi} that after a suitable matching 
of coordinate choices, the central charges and conformal weights 
exactly coincide, providing one piece of evidence for the proposed 
explanation of universality.  There is also an intriguing link 
to the loop quantum gravity approach of section \ref{carLoop}: 
the induced horizon Chern-Simons theory in loop quantum gravity 
is naturally associated with a two-dimensional conformal field 
theory \cite{carWittenb}, whose central charge matches the horizon 
central charge of section \ref{carHorizoncon}.  Searches for 
hidden conformal symmetry in loop quantum gravity, the fuzzball 
approach, and induced gravity are currently underway.

\subsection{What are the states? \label{carStates}}

\index{black hole!microstates!effective description}%
In light of the problem of universality, is there anything 
general we can say about the states responsible for black hole 
thermodynamics?  At first sight, the answer must be ``no'': if 
a universal underlying structure controls the density of states, 
there should be many different models with different degrees of 
freedom but with the same thermodynamic properties.  Nevertheless, 
it may still be possible to find an \emph{effective} description 
that is valid across models.

\index{BTZ black hole}%
\index{black hole!BTZ}%
To see this, let us first return to the BTZ black hole.  In three
spacetime dimensions, general relativity has a peculiar feature: 
it is a topological theory, with no propagating degrees of freedom 
\cite{carCarlipj}.  Where, then, do the black hole degrees of 
freedom come from?  

The answer to this paradox is at least partially understood 
\cite{carCarlipc}.  For the (2+1)-dimensional Einstein-Hilbert 
action to have any black hole extrema, one must impose anti-de
Sitter boundary conditions at infinity.  Diffeomorphisms that do 
not respect these boundary conditions are no longer true invariances 
of the theory, and states one might naively take to be physically 
equivalent---states that differ only by a diffeomorphism---must 
be considered distinct if the diffeomorphism connecting them is 
incompatible with the boundary conditions.  New physical degrees 
of freedom thus appear, which can be labeled by diffeomorphisms 
that fail to respect the anti-de Sitter boundary conditions.  The 
action for these new degrees of freedom can be extracted explicitly 
from the Einstein-Hilbert action \cite{carCarlipk}, and the resulting 
dynamics is that of a Liouville theory, a two-dimensional conformal 
field theory whose central charge matches the classical value 
obtained by Brown and Henneaux \cite{carBrownHen}.  Whether 
one can actually count the states in this theory to reproduce 
the Bekenstein-Hawking entropy remains an open question 
\cite{carCarlipc,carChen}.

\index{horizon constraints}%
For higher dimensional black holes, the problem is quite a 
bit more difficult.  One possible approach is to start with the 
Virasoro algebra (\ref{carCardyform1}) for the near-horizon 
conformal algebra of section \ref{carHorizoncon}.  In Dirac 
quantization, the existence of a constraint ordinarily restricts 
the physical states: we should require that
\begin{equation}
L[\xi]|\mathit{phys}\rangle 
  = {\bar L}[{\bar\xi}]|\mathit{phys}\rangle = 0 .
\label{carWhatstates1}
\end{equation}
But if the central charge $c$ is nonzero, these conditions are 
incompatible with the algebra (\ref{carCardyform1}).  The solution 
is known in conformal field theory---one can, for instance, require 
only that the positive frequency parts of the Virasoro generators 
annihilate physical states \cite{carCFT}---but the result is 
much the same as for the BTZ black hole: certain states that were 
originally counted as nonphysical have now become physical.  While 
it is not exactly the same, this phenomenon is reminiscent of the
Goldstone mechanism \cite{carWeinbergb}, in which a spontaneously 
broken symmetry leads to massless excitations in the ``broken'' 
directions.  And like the Goldstone mechanism, it can provide an 
effective description of degrees of freedom that is independent 
of their fundamental physical makeup.

\index{Virasoro algebra}%
One way to see whether this picture makes sense is to examine 
the path integral measure.  The effect of adding a central 
charge to the Virasoro algebra is to make certain constraints 
second class \cite{carDirac,carDiracb}.  The presence of such 
second class constraints leads to a new term in the measure, 
similar to the Faddeev-Popov determinant in quantum field theory 
\cite{carHenneauxTeit}.  Such a determinant can be interpreted 
as a contribution to the phase space volume, or the density of 
states, and might explain the counting of black hole states.  
For the present case, the relevant determinant is of the form
$$ \det \left|-\frac{c}{12}\frac{d^3\ }{dx^3} + \frac{d\ }{dx}L 
 + L\frac{d\ }{dx}\right|^{1/2} \qquad
\hbox{with}\ \ L = L_0  + L_1e^{2ix} + L_{-1}e^{-2ix} .$$
Work on evaluating and understanding this expression is in progress.

Perhaps the most important test of this idea would be to couple the
effective horizon degrees of freedom to external matter and see if
one could reproduce Hawking radiation.  In 2+1 dimensions, this can
be done \cite{carEmparan}.  In higher dimensions, it may be possible
to take advantage of the conformal description of Hawking radiation 
discussed in section \ref{carStress}, but this remains to be seen.

\section{Open Questions}

Some thirty-five years after the seminal papers of Hawking and 
Bekenstein, black hole equilibrium thermodynamics is a mature 
subject.  The role of trans-Planckian excitations near the 
horizon, discussed in section \ref{carBogol}, is not yet fully 
understood, and questions of possible observational tests remain 
of great interest, but I will risk the claim that the macroscopic 
thermodynamic properties of black holes are largely under control.

The microscopic, statistical mechanical, picture of the black hole, 
in contrast, is poorly understood, and is the subject of a great 
deal of research.  This is hardly surprising---black hole microstates 
are almost certainly quantum gravitational, and we are still far 
from a complete, compelling theory of quantum gravity.

Much of the current research focuses on particular microscopic 
models of black holes, from string theory, loop quantum gravity, 
and a number of other perspectives.  But there are also some
broader open questions.  In these lectures, I have emphasized one
of these, the problem of universality, mainly because it is a focus 
of my own research.  But I will close by briefly mentioning two 
other deep questions.

\subsection{The information loss paradox}

\index{information loss paradox}%
\index{black holes!unitarity}%
Consider a configuration of matter in a pure state---a spherically
symmetric state of a scalar field, for instance---that collapses 
to form a black hole, which then evaporates by Hawking radiation.  
If Hawking radiation is exactly thermal, and if the black hole
evaporates completely, the ultimate result will be a transition 
from an initial pure state to a final mixed (thermal) state 
\cite{carHawkingd}.  Such an evolution is not unitary, and seems 
to violate the basic principles of quantum mechanics.  Similarly, 
we can imagine a black hole held at equilibrium by the continual 
ingestion of mass to balance its Hawking radiation; this would 
seem to allow us to convert an arbitrarily large amount of matter 
from a pure to a mixed state.

The solution to this paradox is heavily debated \cite{carPreskill,%
carBanks,cartHooftb}.  If the black hole horizon is fundamental 
(as it is not in, for instance, the ``fuzzball'' proposal discussed 
in Mathur's lectures \cite{carMathur0}), there is wide agreement 
that any answer must involve a breakdown of locality; see, for 
example, \cite{carGiddingsb,carBalasubramanian,carAshtekarTav}.  
But there is certainly no consensus as to how such a breakdown 
might occur.  The answer is likely to involve deep problems of 
quantum gravity, a setting in which nonlocality is both inevitable 
and very poorly understood \cite{carCarliprev}.

\subsection{Holography \label{carHolography}}

\index{holography}%
As a count of microscopic degrees of freedom, the Bekenstein-Hawking
entropy (\ref{carintro1}) has a peculiar feature: the number of 
degrees of freedom is determined by the area of a surface rather 
than the volume it encloses.  This is very different from 
conventional thermodynamics, in which entropy is an extensive 
quantity, and it implies that the number of degrees of freedom grows 
much more slowly with size than one would expect in an ordinary 
thermodynamic system.

This ``holographic'' behavior \cite{cartHooftc,carSusskindc} seems
fundamental to black hole statistical mechanics, and it has been
conjectured that it is a general property of quantum gravity.  
\index{generalized second law of thermodynamics}%
\index{second law of thermodynamics}%
It may be that the generalized second law of thermodynamics requires
a similar bound for any matter that can be dropped into a black
hole; a nice review of such entropy bounds can be found in
\cite{carBousso}.  The AdS/CFT correspondence discussed in section
\ref{carAdSCFT} is perhaps the cleanest realization of holography in
quantum gravity, but it requires specific boundary conditions.  A
more general formulation proposed by Bousso \cite{carBoussob} is
supported by classical computations \cite{carFlanagan}, and is
currently a very active subject of research, extending far beyond
its birthplace in black hole physics to cosmology, string theory,
and quantum gravity.

\begin{acknowledgement}
These lectures were given during an appointment to the Kramers
Chair at Utrecht University, for whose hospitality I am very
grateful.  This work was supported in part by U.S.\ Department 
of Energy grant DE-FG02-91ER40674.
\end{acknowledgement}

\section*{Appendix: Black Hole Basics}
\addcontentsline{toc}{section}{Appendix}

\index{event horizon}%
Intuitively, a black hole is a ``region of no return,'' an area of
spacetime from which not even light can escape.  For a spacetime that
looks asymptotically close enough to Minkowski space, this intuitive 
picture is formalized by the notion of an event horizon, the boundary 
of the past of future null infinity, that is, the boundary beyond 
which no light ray can reach infinity \cite{carHawkingEllis}.  The 
event horizon has been extensively studied, and has many interesting
global properties: for example, it cannot bifurcate and cannot 
decrease in area.

Unfortunately, while the event horizon has nice properties, it 
does not seem to be quite the right object to capture local physics.  
\index{event horizon!teleological nature}%
The problem is that the event horizon is teleological: that is, 
its definition requires knowledge of the indefinite future.  To 
illustrate this with a thought experiment, imagine that we are at 
the center of a highly energetic ingoing spherical shell of light,
currently two light years from Earth.  Suppose this shell is so
energetic that it has a Schwarzschild radius of one light year.%
\footnote{This is admittedly not very likely, but note that it
cannot be ruled out observationally: no signal could propagate
faster than such a shell, so we would not know of its existence
until it reached us.}  If I now shine a flashlight into the sky,
one year from now the light will have traveled one light year,  
where it will meet the incoming shell just as the shell reaches 
its own Schwarzschild radius.  At that point, the pulse of light
from the flashlight will be trapped at the horizon of an ordinary 
Schwarzschild black hole, and will be unable to travel any farther 
outward.  In other words, in this scenario we are \emph{now} at 
the event horizon of a black hole, even though we will detect no 
change in our local observations until we are abruptly crushed out 
of existence two years from now.

Since it seems implausible that Hawking radiation ``now'' can 
depend on such future events, the event horizon is probably not 
quite the right object for the study of black hole thermodynamics.  
Over the past few years, a number of attempts have been made to 
suitably ``localize'' the horizon; a nice review can be found in 
\cite{carBoothb}.  

\index{isolated horizon}%
In these lectures, I will mainly use the concept of an ``isolated 
horizon'' \cite{carAshtekarc}, a locally defined surface that seems 
appropriate for equilibrium black hole thermodynamics.  An isolated 
horizon is essentially a null surface whose area remains constant 
in time, as the horizon of a stationary black hole does.  A thought 
experiment may again be helpful.  Imagine a spherical lattice 
studded with equally spaced flashbulbs, set to all go off at the 
same time (as measured in the lattice rest frame).  When the bulbs 
flash, they will emit two spherical shells of light, one ingoing 
and one outgoing.  In ordinary nearly flat spacetime, the area of
the outgoing sphere increases with time.  At the horizon of a
Schwarzschild black hole, on the other hand, it is not hard to 
check that the area of the outgoing sphere remains constant, while
inside the horizon, both spheres decrease in area.\footnote{The
outgoing sphere remains outgoing with respect to the lattice, of 
course; as the lattice collapses, its area decreases even faster 
than that of the outgoing light sphere.}   

To generalize this example, we first define a nonexpanding horizon 
$\cal H$ in a $d$-dimensional spacetime to be a $(d-1)$-dimensional 
submanifold such that 
\cite{carAshtekar,carAshtekarc}
\begin{enumerate}
\item $\cal H$ is null, with null normal $\ell_a$;
\item the expansion of $\cal H$ vanishes: $\vartheta_{(\ell)}
 = q^{ab}\nabla_a\ell_b = 0$, where $q_{ab}$ is the induced metric
 on $\cal H$;
\item $-T^a{}_b\ell^b$ is future-directed and causal.
\end{enumerate}
These conditions imply the existence of a one-form $\omega_a$ such 
that
$$\nabla_a\ell^b = \omega_a\ell^b \quad\hbox{on $\cal H$} .$$
The surface gravity for the normal $\ell^a$ is then defined as
\begin{equation}
\kappa_{(\ell)} = \ell^a\omega_a .
\label{carAppendix1}
\end{equation}
Note, though, that the normal $\ell^a$ is not unique: a null vector 
has no canonical normalization, so if $\ell^a$ is a null normal to 
$\cal H$ and $\varphi$ is an arbitrary function, $e^\varphi\ell^a$ 
is also a null normal to $\cal H$.  We can partially fix this scaling 
ambiguity by demanding further time independence: we define a weakly 
isolated horizon by adding the requirement
\begin{enumerate}{\setcounter{enumi}{3}}
\item ${\cal L}_\ell\omega = 0$ on $\cal H$ ,
\end{enumerate}
where $\cal L$ denotes the Lie derivative.  
\index{black hole mechanics!laws of}%
This constraint implies 
the zeroth law of black hole mechanics, that the surface gravity is 
constant on the horizon.   

Even with this last condition, the null normal $\ell^a$ may be 
rescaled by an arbitrary constant.  Such a rescaling also scales 
the surface gravity, so the numerical value of $\kappa_{(\ell)}$ 
remains undetermined.  This reflects a genuine physical ambiguity in 
the choice of time at the horizon.  Note that the first law of black 
hole mechanics (\ref{carFourLaws1}) requires such an ambiguity: mass
is only defined relative to a choice of time, so for consistency,
rescaling time must also rescale the surface gravity.

For a stationary black hole, $\ell^a$ can be chosen to coincide with 
the Killing vector that generates the horizon, whose normalization 
is fixed at infinity---that is, we can use the global properties of
the solution to adjust clocks at the horizon by comparing them to 
clocks at infinity.  In this case, the isolated horizon coincides 
with the Killing horizon discussed in Gernot Neugebauer's lectures 
\cite{carNeugebauer}.  If, on the other hand, we wish to focus on 
physics only at or very near the horizon, the normalization becomes 
more problematic.  One can use the known properties of exact solutions 
to write an expression for the surface gravity in terms of other 
quantities at the horizon, thereby fixing $\ell^a$ \cite{carAshtekar}, 
but so far the procedure seems somewhat artificial.

\index{black hole mechanics!laws of}%
As noted in section \ref{carFourLaws}, weakly isolated horizons obey 
the four laws of black hole mechanics, the second law in the strong
form that the area, by definition, remains constant.  Generalization 
to dynamical, evolving horizons are also possible, and could provide 
a setting for nonequilibrium black hole thermodynamics; for a recent 
review, see \cite{carKrishnan}.

\end{document}